\documentclass[runningheads]{llncs}

\usepackage{cite}
\usepackage{amsmath,amssymb,amsfonts}
\usepackage{algorithmic}
\usepackage{graphicx}
\usepackage{textcomp}
\usepackage{xcolor}
\usepackage{soul}
\usepackage{caption}
\usepackage{subcaption}

\begin{document}
%

\title{Collective Reasoning for Safe Autonomous Systems\thanks{This paper summarizes the content of an invited talk at the DATE 2023 Initiative on Autonomous System Design, 17-19 April 2023, Antwerp, Belgium.}}


%
\author{Selma Saidi}
%
\authorrunning{S. Saidi}
%
\institute{Chair of Embedded Systems\\ TU Dortmund University\\ Dortmund, Germany \\
\email{selma.saidi@tu-dortmund.de}\\
%
}
\maketitle              
\begin{abstract}
Collaboration in multi-agent autonomous systems is critical to increase performance while ensuring safety. However, due to heterogeneity of their features in, e.g., perception qualities, some autonomous systems have to be considered more trustworthy than others when contributing to collaboratively build  a common environmental model, especially under uncertainty. In this paper, we introduce  the idea of increasing the reliability of autonomous systems by relying on collective intelligence. We borrow concepts from social epistemology to exploit individual characteristics of autonomous systems, and define and formalize at design rules for collective reasoning to achieve collaboratively increased safety, trustworthiness and good decision making.

\end{abstract}
%
%
%
\section{Designing Autonomous Systems as an Open Problem}
Autonomous systems have the ability to interact with the environment and act independently by solving complex tasks without human intervention. While designing systems, that are able to do well-defined (repetitive) tasks, like automata, is not new, the main current challenge with designing autonomous systems is to provide them with the ability to operate correctly and safely in a dynamic and open context. That is performing tasks at operation time in an environment that is uncertain or was not fully known or defined at design time. In such environments, autonomous systems must as well \emph{self-decide}. However, when extended to fields like automated driving, where safety is a major requirement, high levels of trustworthiness and good decision making become a necessity and need to be guaranteed by design for the operational phase.

Despite the advancements in the fields of machine learning and artificial intelligence to support increased levels of automation and 
intelligence\footnote{The term intelligence is used here in the sense of the ability to infer new (functional) behavior.} using a growing number of learning-enabled components, reasoning about safety is still lagging behind. Only traditional methods and mechanisms (like designing for worst-case behavior, FMEA, FTA) to guarantee correctness and safety are currently largely used, and there is a tremendous need to  design approaches to reason about safety at operational time. 
In addition to learning-enabled autonomous functions, embedding functions  responsible of safety assurance is of a great importance to guarantee supervision of functional behavior~\cite{SaidiZDE22a}. The goal is to increase trustworthiness and contribute to good decision making of autonomous systems, especially in the presence of uncertainty in the environment where they operate. 






\section{Assured Autonomy Using Collective Intelligence}
Collective intelligence refers to the collective exploitation of distributed intelligence for a better decision making in groups as opposed to individual decision making. Collective intelligence as defined in~\cite{collective_intelligence_def} is a form of universally distributed intelligence, constantly enhanced, coordinated in real-time, and resulting in the effective mobilization of skills. Collective intelligence therefore builds on the premise that a group is collectively "smarter" than a single individual, which globally leads to better decision making processes. Collective intelligence has been exploited in different domains~\cite{collective_intelligence}. A particular focus, e.g., in social computing is on human-machine interaction where interconnected groups of people and computers are required to collectively perform intelligent actions. 

Collective intelligence through cooperation and collaboration using  consensus has been used in autonomous multi-agent systems~\cite{multi-agent} as well.  Therefore to improve reliability and safety in autonomous systems, knowledge acquisition, and possibly decision making\footnote{We distinguish between collaborative knowledge acquisition (e.g., to build a model of the environment) and collaborative decision making (e.g., path planning and maneuvering) as one can be done collectively and not necessarily the other one.}, should be generated by groups of systems and not by single individual ones. 
Collaborative autonomy as defined by~\cite{Gill11}, however, explicitly acknowledges that systems have different features when they collectively cooperate. This distinction may seem minor but is fundamental when \emph{aggregating} knowledge and actions of autonomous systems using collective reasoning. 







\subsection{Collaborative Autonomy}
Collaborative aspects in autonomy exploiting cooperation between autonomous multi-agent
systems and their interaction to solve tasks and maximize utility have been proposed
in several work such as~\cite{multiagent}. Collaborative autonomy can be defined as the principle
underpinning collaborative intelligence through which individual contributors maintain their
roles and priorities as they apply their unique skills and leadership autonomy in a problem-solving process~\cite{Gill11}. 

Collaborative autonomy therefore recognizes that autonomous
systems (or agents) have different capacities, of for instance, pattern recognition, and a
consensus in reaching collaborative intelligence (as traditionally applied in autonomous multi-agent systems) is not always required.
Cooperative driving currently leverages Vehicle-to-X (i.e., vehicle to vehicle and/or vehicle to
infrastructure) communication technologies aiming to carry out cooperative functionalities, such
as cooperative sensing and cooperative maneuvering\footnote{The first fully automated driving car performing automated valley parking and completely relying on infrastructure has been recently released: https://www.bosch.com/stories/autonomous-parking-in-parking-garages/}. Today, infrastructure to support Device-Edge-Cloud Continuum enables sharing of information between distributed (autonomous)
systems and infrastructure\cite{MundhenkHHZ22,keynote_Date2023}. Broadcasting information from every vehicle using for instance collective awareness messages is thereby enabled. Every vehicle can later make a decision locally or globally, like maneuvering, based on this information. The fact, however, that there is an infrastructure to share information does by no means imply that the issue of safety and correct collective behavior is assured.
There is a lack of structured and systematic processes and methods to collectively reason about the shared information in order to improve safety, trustworthiness and good
decision making using collaboration.

\subsection{Collaborative Trustworthiness and Reliability}

When referring to autonomous systems, the term trustworthiness can be defined considering different parameters~\cite{trustworthiness_as}. 
We refer so far to collaboratively increased trustworthiness and reliability in the following sense: the reliability of an autonomous system increases as its processes for gathering information becomes more accurate,  i.e., lead to a more correct representation of the environment. 
A failure can be defined as an error in an information about the environment that an autonomous
system makes compared to the rest of the group (e.g., due to a misclassification in the perception system).

\begin{figure}
	\centering
  \includegraphics[scale = 0.45]
{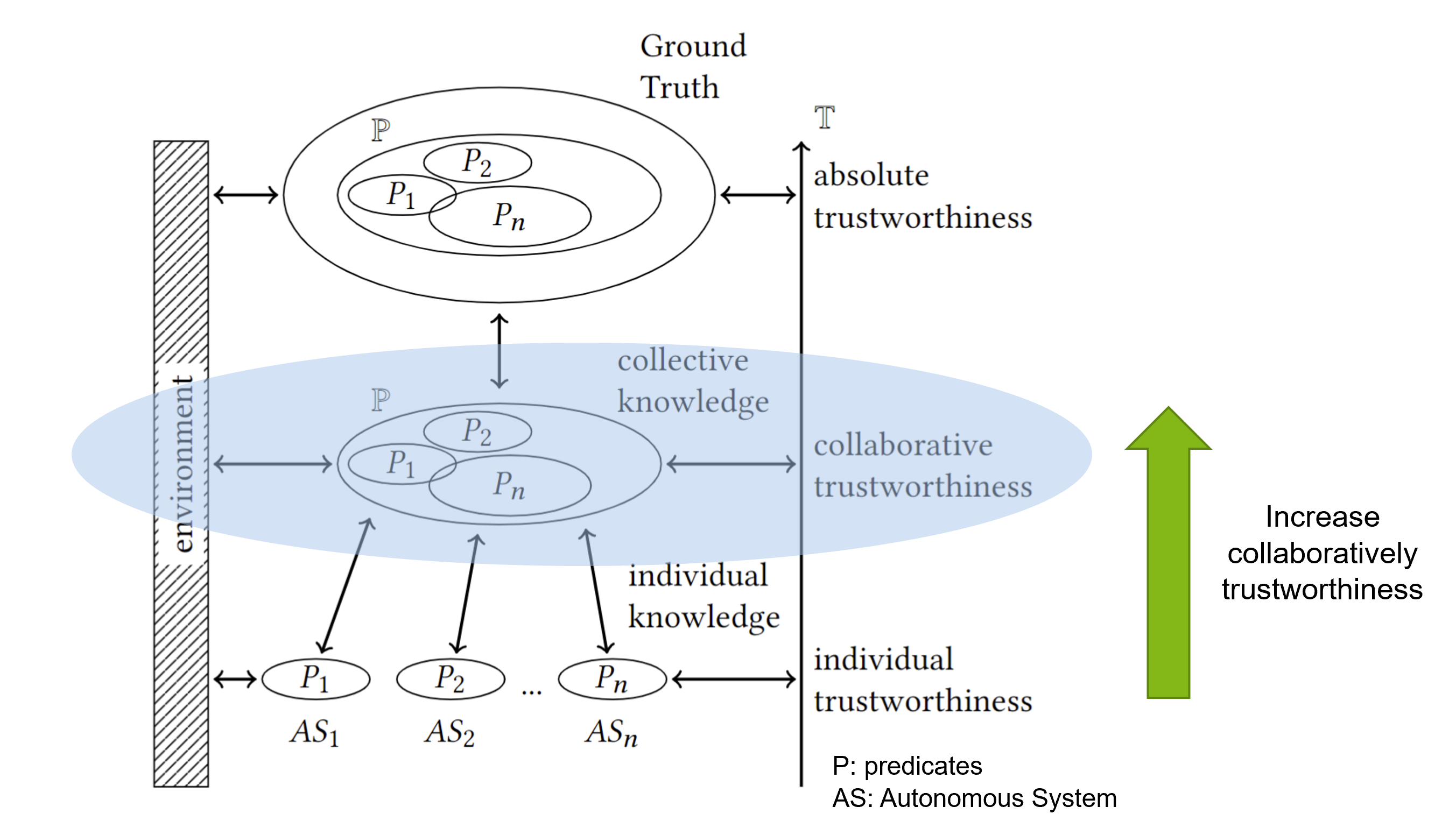}
	\caption{Increase collaboratively trustworthiness exploiting collective knowledge that can be expressed as predicates.}
	\label{fig:collaborative_trustworthiness}
\end{figure}

Instead of individually relying on own individual sphere of knowledge, autonomous systems operating in the same environment can share information about the environment (e.g., in the form of statements like: "an object has been detected" or "distance to this object is less than 500 m", that can be formulated as propositions or predicates). 
 By sharing and exchanging information (totally or partially), the level of trustworthiness in the environment model that leads to good decision making can be collaboratively increased as the sphere of collective knowledge is increase, see Fig~\ref{fig:collaborative_trustworthiness}.   
However more knowledge, aggregated or inferred by the group, does not systematically lead to a better knowledge. 
We need automated collective reasoning methods for defining rules to systematically infer knowledge collectively, as well as decide about the suitable\footnote{Suitable here refers to the group of autonomous systems that are more trustworthy than others in providing better information about the environment.} group of autonomous systems that helps improve collaboratively reliability and  decision making process.

\subsection{Social Epistemology for Collective Reasoning}
\begin{figure}
	\centering
  \includegraphics[scale = 0.55]
{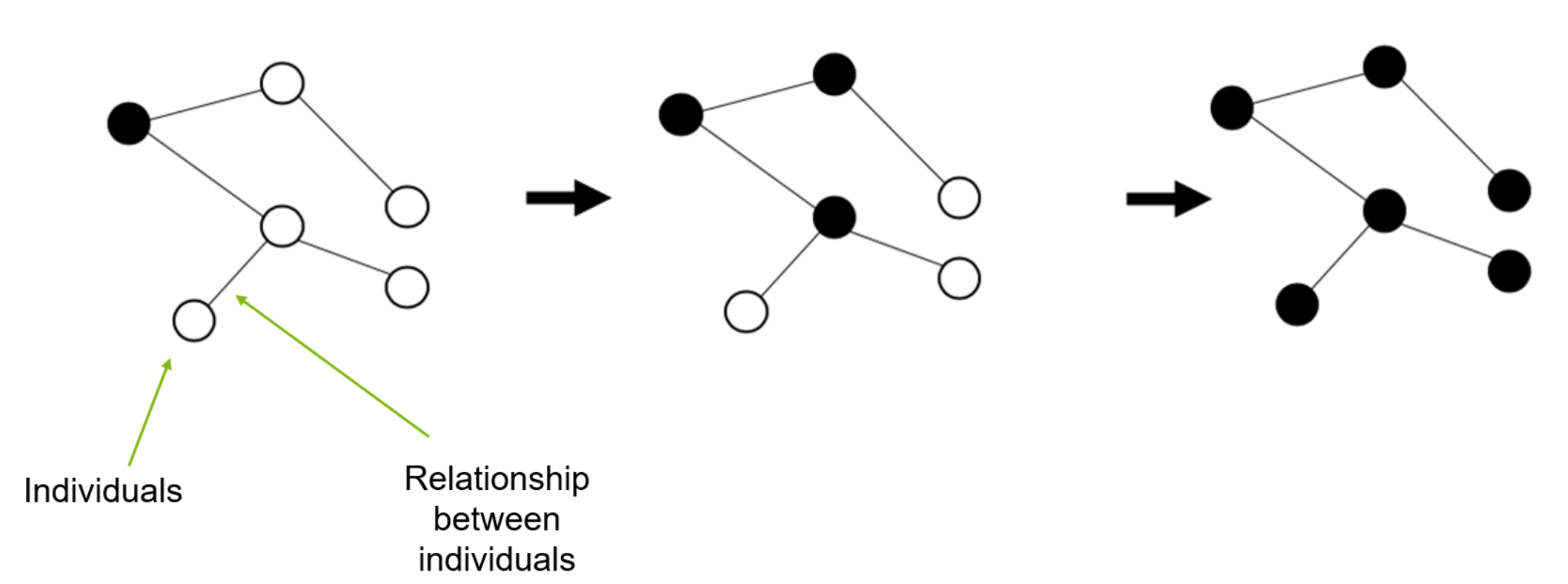}
	\caption{Schematic description based on \cite{sep_epistemology_social} of belief propagation in social epistemology using graphs considering individuals and their relationships.}
	\label{fig: belief-propagation}
\end{figure}

When considering the analogy with humans, reasoning in groups is what humans do regularly in their daily lives to collect more knowledge and improve their decision making process, considering in addition to own opinion, other opinions (often from heterogeneous groups) of individuals with different expertise and different features. Social epistemology and in particular belief propagation deals with the study of how "good" knowledge and belief propagates in a society~\cite{sep_epistemology_social}. This is based on a structuring of the group that can be viewed as a graph, see Fig.,~ \ref{fig: belief-propagation}, where nodes depict individuals in the group and edges are relationships between them. Beliefs can then propagate from one individual to the rest of the group following a set of rules, which determine information and individuals to rely on for  reasoning and how beliefs should be weighted. However, one of the main issues in the epistemic sense is to deal with \emph{peer disagreement}. That is, should an individual (or a system) adjusts an own belief based on additional knowledge, which can as well be contradictory, provided by other peers? In the following, we discuss how similar principles can be applied to autonomous systems.



\section{Autonomous Systems and Rules for Safe Collective Reasoning}

\subsection{Structuring the Group or "Society" of Autonomous Systems}
\begin{figure}
	\centering
  \includegraphics[scale = 0.65]{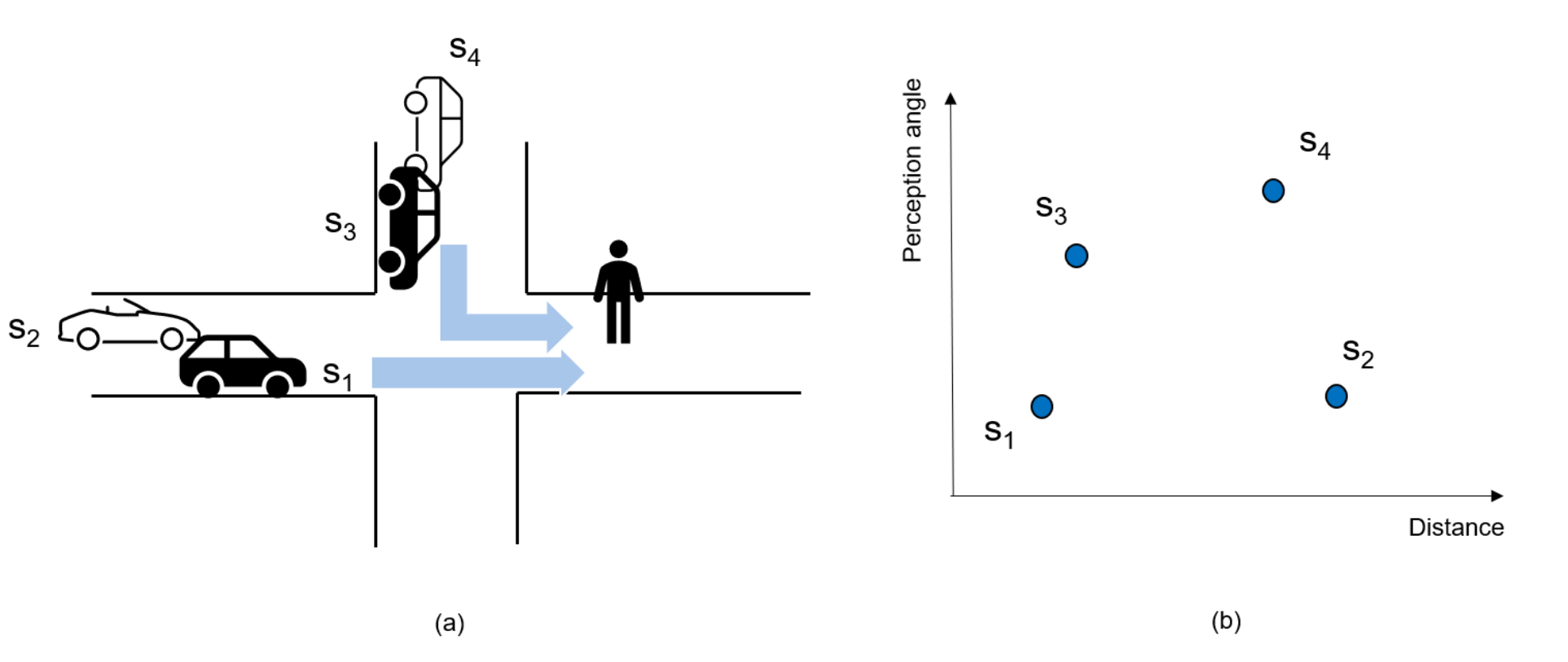} 
	\caption{ Example of autonomous systems operating in the same environment a) vehicles with different features (distance and perception angle) possibly detecting a pedestrian, b) ranking between vehicles based on their features (assuming that the smaller is the value of the feature and the better the quality is).}
	\label{fig:intersection_features}
\end{figure}

\begin{figure}
	\centering
  \includegraphics[scale = 0.55]
{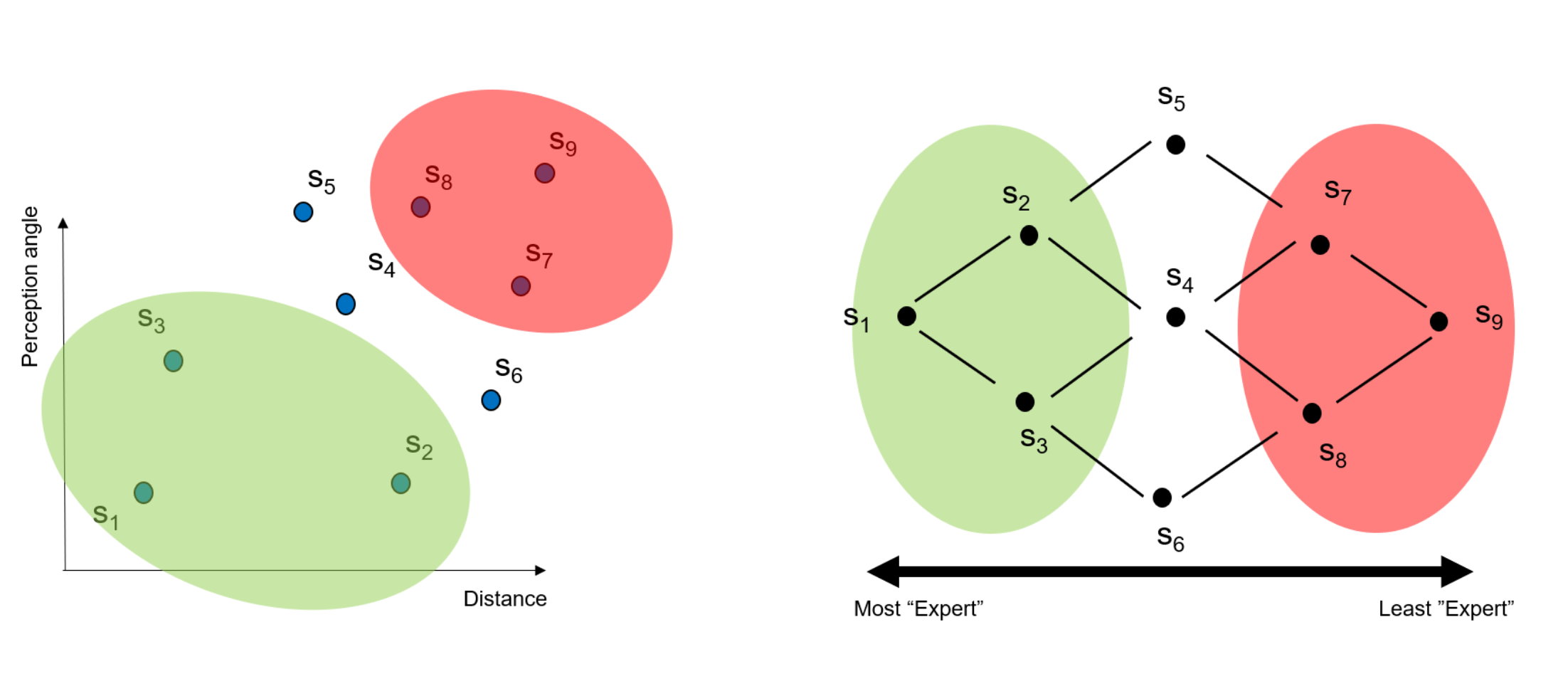}
	\caption{Lattice as a formal representation of partial-order relationship between autonomous systems based on their features. Green (resp. red) area represent autonomous systems with better (resp. worse) quality of features than $s_4$.} 
	\label{fig:lattice}
\end{figure}

By applying the principles of belief propagation in social epistemology, we first need a way to structure the group (or society) of autonomous systems or agents\footnote{In this paper, we use similarly the terms autonomous systems or agent. }. Since autonomous agents are different w.r.t. the quality of their features, we use this criteria as a mean to define relationships between autonomous agents considering \emph{dominance} of the quality of features. Let us consider a simple example of a smart intersection as depicted in Fig.,~\ref{fig:intersection_features} with four vehicles and considering two features: distance and perception angle. It becomes clear from the example that vehicle $s_1$ (resp., $s_4$) has better (resp. worse) distance to the pedestrian and a better (resp. worse) perception angle compared to other vehicles. Both $s_2$ and $s_3$ are not comparable since $s_2$ has a better perception angle but a larger distance to the pedestrian compared to $s_3$, while for $s_3$ it is the opposite. We can therefore provide a ranking of all vehicles present in the intersection based on the quality of their features.  We formulate that structure using a \emph{lattice} as a formal representation for partial ordering, with an infimum and a supremum that constitute autonomous agent with worst and best quality of features respectively, see Fig~\ref{fig:lattice}.

\subsection{Notion of "Expert" in Autonomous Systems}
Considering ranking of autonomous agents based on the quality of their features, a notion of "expertise" can be defined to qualify autonomous agents of a better quality. For every autonomous system with a given quality of features (e.g., $\{s_4\}$ in Fig,~\ref{fig:lattice}), experts are the set of points whose quality dominate its quality for every feature (e.g., $\{s_1, s_2, s_3\}$). Other autonomous agents whose quality of features are dominated by this point can be referred to as less experts. In fact, every autonomous system may still provide a given useful information/knowledge and may have some level of expertise.

\subsection{Example of Simple Rules for Belief Propagation}

In order to illustrate the concept of rules, we consider in the following very simple and straightforward set of rules for belief propagation that can be applied, and discuss that, when applied, these latter do not necessarily lead to a collectively increased reliability.

\begin{figure}
\begin{subfigure}{.33\textwidth}
  \centering
  \includegraphics[width=1\linewidth]{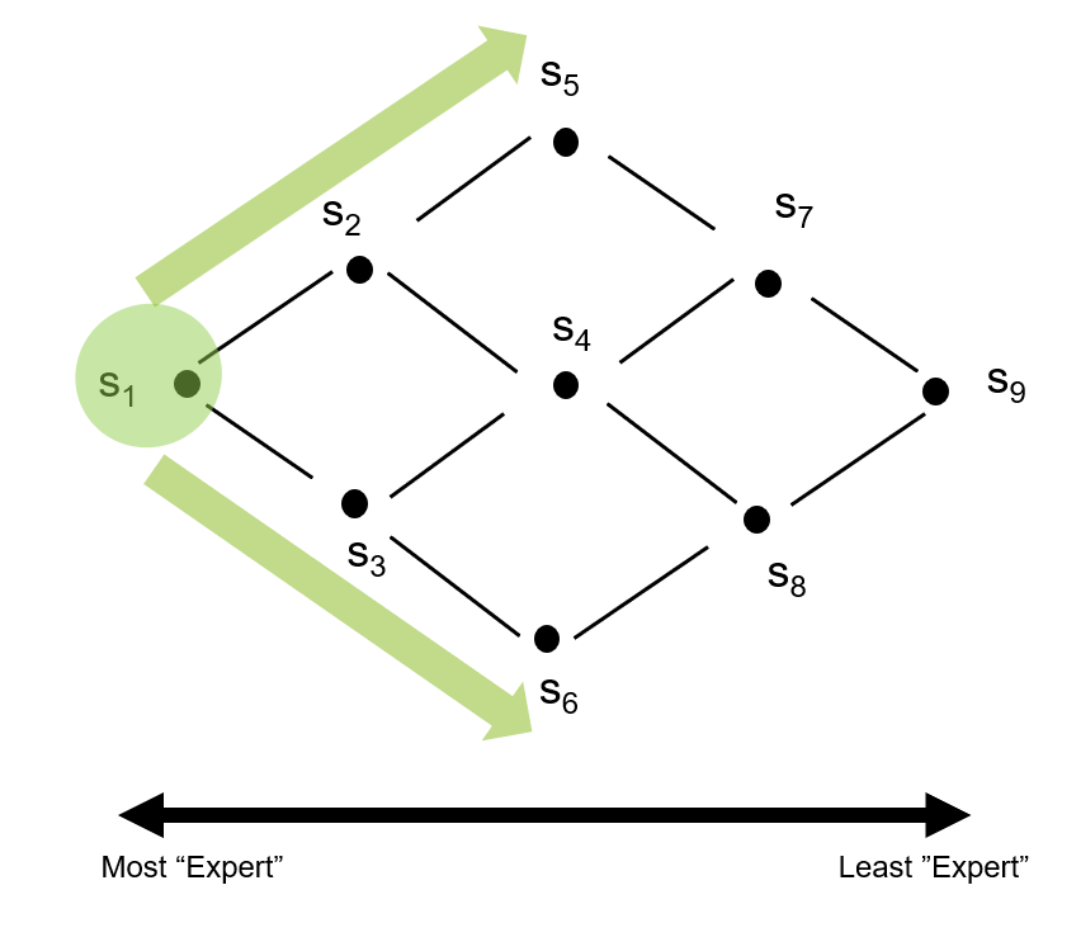}
 \caption{}
  \label{fig:sub1}
\end{subfigure}%
  \hfill
\begin{subfigure}{.33\textwidth}
 \includegraphics[width=1\linewidth]{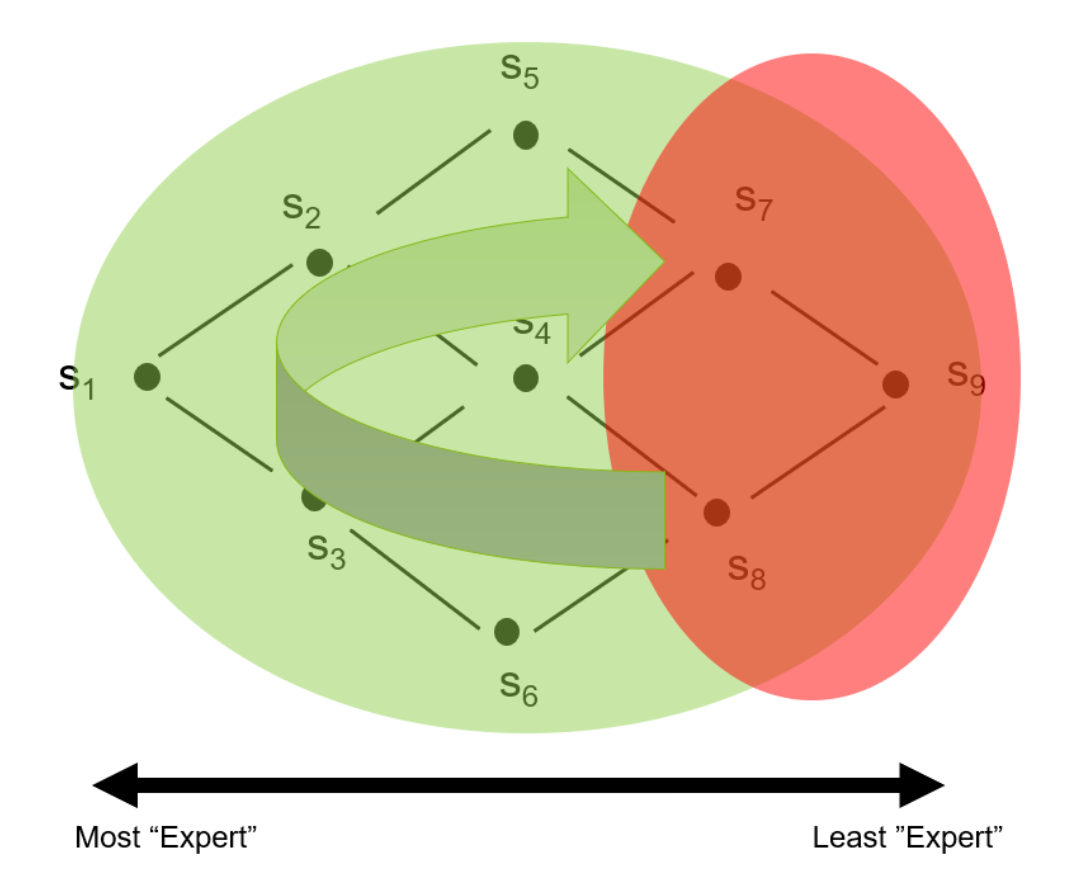}
  \caption{}
 \label{fig:sub2}
\end{subfigure}
  \hfill
\begin{subfigure}{.33\textwidth}
 \includegraphics[width=1\linewidth]{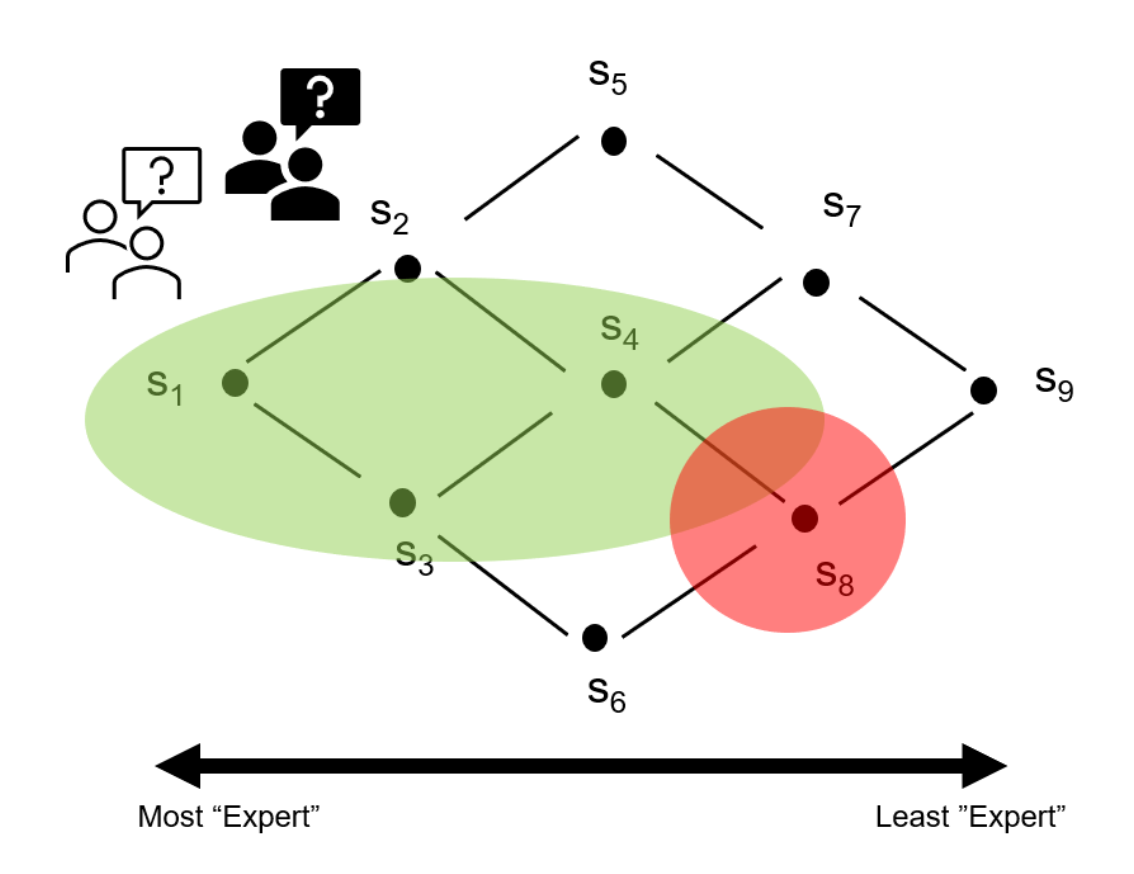}
  \caption{}
 \label{fig:sub3}
 
\end{subfigure}

\caption{Schematic illustration of groups involved in forming a correct belief and its propagation based on different rules: (a)"Most Expert" rule: only the belief of the most expert propagates, (b) "Majority" rule: propagation of all beliefs and the most common belief is then further propagated, (c) further rules involving only a sub-group of autonomous systems with better (and possibly less better) experts.}
\label{fig:rules}
\end{figure}

\paragraph{"Most Expert" rule}
In the most expert rule, the belief (e.g., an object has been detected) of the most "expert" autonomous system is propagated to the rest of the group, see Fig.~\ref{fig:rules}-(a). Each member of the group systematically adopts the belief of the most expert autonomous system, even if contradicting with own belief (e.g., an object has not been detected) or beliefs from other autonomous systems. The rule builds on trusting the opinion of the most expert, 
as it constitutes the autonomous system which has the best quality of features in the group and which thereby considers its belief as the most trustworthy. This rule considers that only the most expert is responsible of forming a correct belief. However, one weakness of the rule is that it relies on \emph{one} single opinion and is therefore not resilient to the case where the most expert autonomous system makes an error of judgement. 

\paragraph{"Majority" rule}
In the majority rule, beliefs from all autonomous system are propagated to the rest of the group. Each autonomous systems therefore collects all beliefs including own belief and adopts the belief of the majority of the group similar to a voting mechanism, see Fig.,~\ref{fig:rules}-(b). In this case, all autonomous systems in the group are responsible of forming a correct belief. While this approach offers a given notion of fairness as all beliefs are considered, this can as well constitute the main weakness of the approach. Indeed, the majority rule considers all beliefs in the group including beliefs from less experts autonomous systems with a low quality of features. In this case all autonomous systems, including less trustworthy ones contribute to forming a correct belief. In the smart intersection example considered previously, even vehicles with a larger distance from the pedestrian or with a bad perception angle will contribute to the overall decision making process, thereby possible leading to an error in the propagated belief and overall decrease in reliability when using the majority rule for collective reasoning.

\section{Challenges to collective reasoning for safe autonomous systems}
In the following we discuss three important challenges in exploiting collective reasoning in collaborative scenarios to improve overall trustworthiness and reliability. These challenges concern the main questions of correctness, efficient computation and design to support autonomous adaptation. 

\paragraph{Define "appropriate" rules for collective reasoning}
As discussed previously, depending on the choice of the applied rule(s) for collective reasoning, they may (or not) lead to a better level of trustworthiness and reliability. Defining suitable\footnote{Note that criteria to define what is suitable need to be precisely defined and their benefit quantified.} rules for collective reasoning involves the following, 
\begin{enumerate}
    \item determining the right (sub-)group of interest that contributes to forming a correct belief, see Fig~\ref{fig:rules}-(c). Note that the larger the considered group of interest, and the higher is the likelihood for peer disagreements. 
    This implies determining the right level of expertise and right features to consider. If only one feature is relevant, then autonomous systems with better quality in that particular feature might be sufficient to consider (e.g.,  a total order of autonomous systems based on a single feature to define levels of expertise for that particular feature). 

    \item 
    When different beliefs from multiple autonomous systems are considered, an aggregate function is needed to infer a correct belief of the group, that can later be propagated. In the majority rule, a max function on the number of similar beliefs is considered. However, determining aggregate functions for inferring "good" knowledge to increase collaboratively trustworthiness and reliability is still an open question, especially that there is no one-size-fits-all aggregate function, several different aggregate functions can be applied depending on the situation or context. 
   Moreover, main requirement on defined rules for aggregation are being i) \emph{deterministic}: applying similar set of rules will lead to the same inferred result. ii) \emph{consistent}: applying two or more consistent rules will not lead to an inferred inconsistent result \cite{impossibility_theorem}.
\end{enumerate}

\paragraph{Complexity and Computation}
One major challenge is the complexity of computation, as the number of considered features increase, the size of the lattice also increases in addition to the number of considered autonomous systems. In the aforementioned example of smart intersection, only two features are considered as distance and perception angle. In practice many other features can be considered to characterize for instance sensors quality. Therefore, when considering $n$ features the number of dimensions in the lattice is also $n$ when modeling all features. Note that all features are not static features, and their quality can therefore change over time. Taking as an example perception angle as a feature, ranking of autonomous systems based on the quality of this feature changes over time. Therefore, the considered lattice as a model for this ranking is not static but rather dynamic and is changing over time. Providing efficient structures for storing and computing such high dimensional lattices for real-time computations is a main challenge. Another challenge is to support efficient updates of such structures for scalability (when new features are considered) and adaptability (i.e., when modifications in the features or their quality need to be considered). 

\paragraph{From Automated to Autonomous Reasoning}
Rules defined for collective reasoning can also change over time, depending on the context of operation the right set of rules that increases trustworthiness maybe different. Furthermore, as autonomous systems evolve, the set of rules that can be applied can as well evolve, new rules for inferring new epistemic knowledge can be generated and autonomously defined by the system. Such step of evolution of the capabilities of collective reasoning can then be labeled as a move from automated  to \emph{autonomous} reasoning. 




\bibliographystyle{plain}
\bibliography{biblio}

\begin{thebibliography}{10}

\bibitem{multiagent}
Louise~A. Dennis and Michael Fisher.
\newblock Verifiable self-aware agent-based autonomous systems.
\newblock {\em Proceedings of the IEEE}, 108(7):1011--1026, 2020.

\bibitem{keynote_Date2023}
Dirk Elias, Dirk Ziegenbein, Philipp Mundhenk, Arne Hamann, and Anthony Rowe.
\newblock The cyber-physical metaverse - where digital twins and humans come
  together.
\newblock In Ian O'Connor, Robert Wille, and Ioana Vatajelu, editors, {\em 2023
  Design, Automation {\&} Test in Europe Conference {\&} Exhibition, {DATE}
  2023, Antwerp, Belgium, April 17-19, 2023}. {IEEE}, 2023.

\bibitem{Gill11}
Zann Gill.
\newblock Collaborative intelligence in living systems: algorithmic
  implications of evo-devo debates.
\newblock In Natalio Krasnogor and Pier~Luca Lanzi, editors, {\em 13th Annual
  Genetic and Evolutionary Computation Conference, {GECCO} 2011, Companion
  Material Proceedings, Dublin, Ireland, July 12-16, 2011}, pages 803--804.
  {ACM}, 2011.

\bibitem{sep_epistemology_social}
Alvin Goldman and Cailin O’Connor.
\newblock {Social Epistemology}.
\newblock In Edward~N. Zalta, editor, {\em The {Stanford} Encyclopedia of
  Philosophy}. Metaphysics Research Lab, Stanford University, {W}inter 2021
  edition, 2021.

\bibitem{trustworthiness_as}
S.~Kate~Devitt.
\newblock {\em Trustworthiness of Autonomous Systems}, pages 161--184.
\newblock Springer International Publishing, Cham, 2018.

\bibitem{collective_intelligence_def}
Pierre Levy and Robert Bononno.
\newblock {\em Collective Intelligence: Mankind's Emerging World in
  Cyberspace}.
\newblock Perseus Books, USA, 1997.

\bibitem{impossibility_theorem}
Christian List and Philip Pettit.
\newblock Aggregating sets of judgments: An impossibility result.
\newblock {\em Economics \&amp; Philosophy}, 18(1):89–110, 2002.

\bibitem{collective_intelligence}
Thomas~W. Malone and Michael~S. Bernstein.
\newblock {\em Handbook of Collective Intelligence}.
\newblock The MIT Press, 2015.

\bibitem{MundhenkHHZ22}
Philipp Mundhenk, Arne Hamann, Andreas Heyl, and Dirk Ziegenbein.
\newblock Reliable distributed systems.
\newblock In Cristiana Bolchini, Ingrid Verbauwhede, and Ioana Vatajelu,
  editors, {\em 2022 Design, Automation {\&} Test in Europe Conference {\&}
  Exhibition, {DATE} 2022, Antwerp, Belgium, March 14-23, 2022}, pages
  287--291. {IEEE}, 2022.

\bibitem{SaidiZDE22a}
Selma Saidi, Dirk Ziegenbein, Jyotirmoy~V. Deshmukh, and Rolf Ernst.
\newblock Autonomous systems design: Charting a new discipline.
\newblock {\em {IEEE} Des. Test}, 39(1):8--23, 2022.

\bibitem{multi-agent}
Michael Wooldridge.
\newblock {\em An Introduction to Multiagent Systems}.
\newblock Wiley, Chichester, UK, 2 edition, 2009.

\end{thebibliography}

\end{document}